       \newcommand{\beq}{\begin{equation}}
       \newcommand{\eeq}{\end{equation}}
       \newcommand{\bea}{\begin{eqnarray}}
       \newcommand{\eea}{\end{eqnarray}}
       \newcommand{\beas}{\begin{eqnarray*}}
       \newcommand{\eeas}{\end{eqnarray*}}
       
       \newcommand{\bu}{{\mathbf u}}

       \newcommand{\bE}{{\mathbf E}}

\newcommand{\nonu}{\nonumber \\}

\documentclass[12pt]{article}
\usepackage[dvips]{graphicx}\usepackage[dvips]{graphics}
\usepackage{enumerate}
\usepackage{psfrag}
\usepackage{color}
\begin{document}
\begin{center}
{\Large \bf Tokamak-like Vlasov equilibria } \vspace{3mm}

{\large H. Tasso$^1$, G. N. Throumoulopoulos$^2$}
%, H. Tasso$^3$} \vspace{2mm}

{\it $^1$Max-Planck-Institut f\"{u}r Plasmaphysik,  
Euratom Association, \\  85748 Garching bei M\"{u}nchen, Germany}
\vspace{2mm}

 { \it  $^2$Department of  Physics, University of Ioannina, \\ Association Euratom-Hellenic Republic, 
 GR 451 10 Ioannina, Greece} \vspace{2mm}

%{\it $^3$Max-Planck-Institut f\"{u}r Plasmaphysik,  
%Euratom Association, \\  85748 Garching bei M\"{u}nchen, Germany} \vspace{2mm}

Emails:het@ipp.mpg.de, \  gthroum@cc.uoi.gr
%, \  het@ipp.mpg.de
\end{center}
%\noindent
%
%
%\vspace{2mm}
%\begin{center}
%{\large \it December 2000}
%\end{center}
\vspace{2mm}
\begin{center}
{\bf \large Abstract}
\end{center}
 \noindent

 Vlasov equilibria of axisymmetric plasmas with vacuum toroidal magnetic field can be reduced, up to a selection of ions and electrons distributions functions, to a Grad-Shafranov-like equation. Quasineutrality narrow the choice of the distributions functions. In contrast to two-dimensional translationally symmetric equilibria whose electron distribution function consists of a displaced Maxwellian, the toroidal equilibria need deformed Maxwellians. In order to be able to carry through the calculations, this deformation is produced by means of either a Heaviside step function or an exponential function. The resulting Grad-Shafranov-like equations are established explicitly.
\vspace{1mm}


\newpage

\vspace{20mm}
\section{Introduction}
In a previous paper \cite{tat1}, it has been proved that the current on the magnetic axis of an axisymmetric Vlasov equilibrium vanishes if the gradient of the distribution function and the electric field are taken equal to zero. However, for a translation symmetric two-dimensional configuration, quasineutral equilibria with non-vanishing current density were explicitly found in \cite{sch,tht1}. In the present contribution, we consider a toroidal configuration with a finite gradient of the distribution function on the magnetic axis and either  vanishing or not vanishing electric field on that axis. We show that the case of \cite{sch,tht1} can be extended to the toroidal case up to the solution of a Grad-Shafranov-like equation with a transcendental RHS.

In section 2, we examine the constants of motion and the related distribution functions. Section 3 is devoted to quasineutrality with zero electric field on axis and section 4  establishes in this case a static Grad-Shafranov-like equation by making a Heaviside-function deformation of the electron distribution function. In section 6 we choose exponentially  deformed ion and electron distribution functions with finite electric fields on axis to derive  a stationary  Grad-Shafranov-like equation with toroidal ion fluid flow.  Section  7 summarises the conclusions.

\section{Constants of motion and distribution functions}

In axisymmetric torus only two constants of motion are known, the energy $E$ and the angular momentum C. For the ions we have:  

\begin{equation}
E_i = e \Phi (r,z) + \frac{M}{2} (v_r^{2} + v_\phi ^{2} + v_z ^{2}),
\end{equation}

\begin{equation}
C_i = M (rv_\phi + erA_\phi).
\end{equation}

For the electrons

\begin{equation}
E_e = - e \Phi (r,z) +\frac{m}{2} (v_r^{2} + v_\phi ^{2} + v_z ^{2}),
\end{equation}

\begin{equation}
C_e = m (rv_\phi - erA_\phi),
\end{equation}
where $\Phi$ is the electrostatic potential, $A_\phi$ the toroidal component of the vector potential, $M$ and $m$ are the masses of ions and electrons respectively. The system $r$, $\phi$, $z$ is the cylindrical system of coordinates and $v_r$, $v_\phi$ $v_z$ are the components of the particle velocities along that system. The charge $e$ is taken as the absolute value of the electron charge. 
 
The solutions of the ion and electron Vlasov equations are given as
\begin{equation}
f_i = f_i (E_i, C_i),
\end{equation}
\begin{equation}
f_e = f_e (E_e, C_e),
\end{equation}
with a normalization of $f_i$ and $f_e$ equal to the total number of particules $N$ so that the densities are given by
\begin{equation}
n_i = \int f_i (E_i, C_i)d^3v,
\end{equation}
\begin{equation}
n_e = \int f_e (E_e, C_e)d^3v.
\end{equation}
The electrical current density $j_\phi$ is 
\begin{equation}
j_\phi= e\int v_\phi f_i (E_i, C_i)d^3v - e\int v_\phi f_e (E_e, C_e)d^3v
\end{equation}

\section{Quasineutrality and electric field on axis}

We assume quasineutrality through the whole plasma instead of Poisson equation for the electric potential. Also in this and the following two sections we  demand that the electric field vanishes on axis.
% to make sure that no ${\bf E}\times{\bf B}$ drift occurs on the magnetic axis.
 A similar condition of vanishing electric field on axis was adopted in Ref.[1] for a near axis consideration of the Vlasov equation.  This leads to 

\begin{equation}
n_i = n_e
\end{equation}
everywhere in the plasma and,  in particular 
\begin{equation}
\nabla n_i = \nabla n_e
\end{equation}
on the magnetic axis.

We introduce now $\Psi = rA_\phi$ as the poloidal flux around the magnetic axis, a function which labels the magnetic surfaces and can be taken equal to zero on the magnetic axis. Also, we compute explicitly both members of equations (10) and (11) using (7) and (8) to obtain

\begin{equation} 
\int f_i (E_i, C_i)d^3v = \int f_e (E_e, C_e)d^3v,
\end{equation}

\begin{equation}
\nabla n_i = \int\left[e\frac{\partial f_i}{\partial E_i}\nabla \Phi + eM\frac{\partial f_i}{\partial C_i}\nabla \Psi + Mv_\phi \frac{\partial f_i}{\partial C_i} \nabla r\right] d^3v,  
\end{equation}
\begin{equation}
\nabla n_e = \int\left[-e\frac{\partial f_e}{\partial E_e}\nabla \Phi - em \frac{\partial f_e}{\partial C_e}\nabla \Psi + mv_\phi \frac{\partial f_e}{\partial C_e} \nabla r\right] d^3v.
\end{equation}

We see from Eq.(12) that the electrostatic potential will be, in general, a function of $\Psi$ and $r$ in contrast with the scale factor free two-dimensional case treated in \cite{sch,tht1}. Similarly, equations (13) and (14) show that the gradient of the electrostatic potential does not necessarily vanish at the magnetic axis given by $\nabla \Psi = 0$ because $\frac{\partial f_i}{\partial C_i}$ and $\frac{\partial f_e}{\partial C_e}$ cannot vanish unless the toroidal current vanishes also. This leads us to look for special choices for $f_i$ and $f_e$ which allow us to have $j_\phi$ different from zero, but, at the same time, have $\nabla \Phi = \nabla \Psi = 0$ on the magnetic axis. This will be treated in the next section.    
 
\section{Static Grad-Shafranov-like equation}

In order to fulfil Eqs. (11), (13) and (14) on the magnetic axis, as just mentioned, we assume for simplicity, $\frac{\partial f_i}{\partial C_i} = 0$ and $\frac{\partial f_e}{\partial C_e}$ symmetric in $v_\phi$ on the magnetic axis i.e. for $\Psi = 0$. Specifically, and for reason of tractability, we choose
\begin{equation}
f_i = \exp\left\{-\beta_i\left[e\Phi + \frac{M}{2}(v_r^2 + v_\phi^2
 + v_z^2)\right]\right\},
\end{equation}

\begin{equation}
f_e = \left[1 + \alpha H(C_e)\right]\exp\left\{\beta_e\left[e\Phi - \frac{m}{2}(v_r^2 + v_\phi^2 + v_z^2)\right]\right\} ,
\end{equation}
where $H(C_e)$ is the Heaviside step function (see e.g. \cite{wei}) and $0 < \alpha \leq 1$. Though physically, singular functions like (16) are not desirable, they are very convenient for the analysis. 
%See also item d) of the Conclusions.

Inserting (15) in (7) we have 
\begin{equation}
n_i = \left(\sqrt{\frac{2\pi}{M}}\right)^3 \exp(-e\beta_i\Phi).
\end{equation}
Let us now consider the impact of the $\alpha$ term of (16) on $n_e$ of Equation (8). Consider the integral over $v_\phi $ 
\begin{equation}
\int_{-\infty}^{\infty} H\left[m(rv_\phi - e\Psi)\right]\exp\left[-\frac{m\beta_e}{2}v_\phi^2\right] dv_\phi
\end{equation}
or
\begin{equation}
\int_{\frac{e\Psi}{r}}^{\infty} \exp{(-\frac{m\beta_e}{2}v_\phi^2}) dv_\phi.
\end{equation}
Introduce $t = \sqrt{\frac{m\beta_e}{2}}v_\phi$ and the "complementary error function" (see \cite{wei}) 
\begin{equation}
erfc(x) = \frac{2}{\sqrt{\pi}}\int_{x}^{\infty}\exp{(-t^2})dt,
\end{equation}
then (19) becomes 
\begin{equation}
\frac{\sqrt{\pi}}{2}\sqrt{\frac{m\beta_e}{2}}erfc\left(\sqrt{\frac{m\beta_e}{2}}\frac{e\Psi}{r}\right).
\end{equation}
Inserting (16) and (21) in (8) we obtain
\begin{equation}
n_e = \left[\left(\sqrt{\frac{2\pi}{m}}\right)^3 + \alpha\frac{\sqrt{\pi}}{2}
\left(\sqrt{\frac{2\pi}{m}}\right)^2\sqrt{\frac{m\beta_e}{2}}erfc\left(\sqrt{\frac{m\beta_e}{2}}\frac{e\Psi}{r}\right)\right]\exp{(e\beta_e\Phi)}.
\end{equation}
Equating (17) and (22) leads to
\begin{equation}
e\beta_e\Phi = - \frac{\beta_e}{\beta_e + \beta_i} log\left[\left(\frac{M}{m}\right)^\frac{3}{2} + \alpha \left(\frac{M}{m}\right)\sqrt{\frac{M}{2\pi}}
\frac{\sqrt{\pi}}{2}\sqrt{\frac{m\beta_e}{2}}erfc\left(\sqrt{\frac{m\beta_e}{2}}\frac{e\Psi}{r}\right)\right]
\end{equation}
or
\begin{equation}
\exp{(e\beta_e\Phi)} = \exp\left\{-\left[\left(\frac{M}{m}\right)^\frac{3}{2} + \alpha \left(\frac{M}{m}\right)\sqrt{\frac{M}{2\pi}}
\frac{\sqrt{\pi}}{2}\sqrt{\frac{m_e\beta_e}{2}}erfc\left(\sqrt{\frac{m_e\beta_e}{2}}\frac{e\Psi}{r}\right)\right]^\frac{\beta_e}{\beta_e + \beta_i}\right\}.
\end{equation}

To compute $j_\phi$ from Eq.(9) we calculate first 
\begin{equation}
-e\int_{\frac{e\Psi}{r}}^{\infty}v_\phi\exp{\left(-\frac{m\beta_e}{2}v_\phi^2\right)} H\left[m(rv_\phi - e\Psi)\right]dv_\phi,
\end{equation}
which is equal to
\begin{equation}
e\int_{\frac{e^2\Psi^2}{2r^2}}^{\infty}\exp{(-xm\beta_e)} dx = -\frac{e}{m\beta_e}\exp{\left(-\frac{e^2\Psi^2}{2r^2}\right)}
\end{equation}
after changing to $x = \frac{v_\phi^2}{2}$.
Using (9), (16) and (26) we obtain
\begin{equation}
j_\phi = -e\alpha\exp{(\beta_e\Phi)}\left(\sqrt{\frac{2\pi}{m}}\right)^2\frac{1}{m\beta_e}\exp{\left(-\frac{e^2\Psi^2}{2r^2}\right)},
\end{equation}
where $\exp{(e\beta_e\Phi)}$ is given by (24).

${\bf B}$ can be written as
\begin{equation}
{\bf B} = \frac{B_0}{r}{\bf e_\phi} +\nabla\Psi \times \frac{{\bf e_\phi}}{r},
\end{equation}
where $B_0$ is the magnitude of a vacuum toroidal field at some value of $r$. Projecting the curl of ${\bf B}$ on ${\bf e_\phi}$ and equating it to (27), we obtain 
\begin{equation}
\frac{1}{r}\left[\frac{\partial^2 \Psi}{\partial r^2} - \frac{1}{r}\frac{\partial \Psi}{\partial r} + \frac{\partial^2 \Psi}{\partial z^2}\right] =  e\alpha\exp{(e\beta_e\Phi)}\frac{2\pi\beta_e}{m^2} \exp{\left(-\frac{e^2\Psi^2}{2r^2}\right)}.
\end{equation}

Eq.(29) is a Grad-Shafranov-like nonlinear elliptic equation whose solution can be found numerically as e.g. a Dirichlet boundary value problem. In particular, the location of the magnetic axis is given by $\Psi = \nabla \Psi = \nabla \Phi = 0$. Since the functions $\exp{(e\beta_e\Phi)}$ and $\exp{\left(-\frac{e^2\Psi^2}{2r^2}\right)}$ are monotonically decreasing functions when $\Psi$ increases from zero to larger values, it can be proved (see \cite{chi}) that, for a given $\Psi$ at the boundary of the relevant domain, a unique solution exists. 

%\newpage

Since the right hand sides of equations (24) and (29) do not depend upon $r$ if $\Psi$ is replaced by $rA_\phi$, it may be advantageous to write (29) in terms of $A_\phi$. This can be easily achieved by inserting $\Psi = rA_\phi$ in (29) to obtain

\begin{equation}
\left[\frac{\partial^2 A_\phi}{\partial r^2} + \frac{1}{r}\frac{\partial A_\phi}{\partial r} + \frac{\partial^2 A_\phi}{\partial z^2} - \frac{A_\phi}{r^2}\right] =  e\alpha\exp{(e\beta_e\Phi)}\frac{2\pi\beta_e}{m^2} \exp{\left(-\frac{e^2 A_\phi^2}{2}\right)}
\end{equation}
where $\exp{(e\beta_e\Phi)}$ from (24) is now a function of $A_\phi$ only without an $r$ dependence.

\section{Higher moments}
It is possible to calculate all the moments of the electron distribution function given by Eq.(16). The moments reduce essentially to the form
\begin{equation}
\int_{-\infty}^{\infty}\exp{(-ax^2)}x^n H\left[(rx - e\Psi)\right]dx
\end{equation}
with $a = \frac{m\beta_e}{2}$.
For n odd we introduce $y = x^2$ so that (31) becomes 
\begin{equation}
\int_{\frac{e^2\Psi^2}{r^2}}^{\infty} \exp{(-ay)} y^\frac{n-1}{2} dy.
\end{equation}
The well known recursion formula
\begin{equation}
\int\exp{(ay)}y^mdy = \frac{1}{a}y^m\exp{(ay)} - \frac{m}{2a}\int y^{(m - 1)} \exp{(ay)} dy
\end{equation}
allows us to reduce the integral (32) to powers of $y$ and $\exp{(-ay)}$.

For n even we can repeatedly take the derivatives of (19) with respect to $a$ to obtain (31). Those derivatives can be expressed in terms of the erfc function and the Hermite polynnomials as can be found in Ref.\cite{wei}. 

\section{Grad-Shafranov-like equation with flow}

To consider equilibria with macroscopic plasma (ion) flow we make now the following choices of the distribution functions
\beq
\label{iondf}
f_i=n_{i0}\left(M\frac{\beta_i}{2\pi}\right)^{3/2}\exp(-\beta_i E_i)\exp(\beta_i V_{i \phi} C_i/r_0), 
\eeq
\beq
\label{elecdf}
f_e=n_{e0}\left(m\frac{\beta_e}{2\pi}\right)^{3/2}\exp(-\beta_e E_e)
\exp(\beta_e V_{e \phi} C_e/r_0), 
\eeq
with $n_{i0}$, $n_{e0}$ $r_0$, $V_{i \phi}$ and $V_{e \phi}$ constant quantities; $E_i$ and $E_e$ as given by (1) and (3);  and 
$$ C_i=M(r v_\phi +e \Psi),    \ \   C_e=m(r v_\phi -e \Psi).$$
Using (7) and (8) one finds for the densities 
\beq
\label{deni}
n_i=n_{i0}\exp\left[\frac{M V_{i \phi}(2 e r_0 \Psi+r^2 V_{i \phi})\beta_i}{2 r_0^2}\right]\exp(-e \beta_i \Phi) 
\eeq
\beq
\label{dene}
n_e=n_{e0}\exp\left[\frac{m V_{e \phi} (-2 e r_0 \Psi+r^2 V_{e \phi})\beta_i}{2 r_0^2}\right]\exp(e \beta_e \Phi).
\eeq
Then on the basis of the quasineutrality condition, $n_i=n_e$, the electrostatic potential can be expressed in terms of $\Psi$ and $r$ as
\bea
\label{Phi}
\Phi(\Psi,r)&=&\left[e(\beta_i+\beta_e)\right]^{-1} \nonu 
 & & log\left\{\frac{n_{i0}}{n_{e0}}
\exp\left[\frac{2 e \Psi r_0 (m V_{e \phi} \beta_e+M V_{i \phi} \beta_i)+r^2(M V_{i \phi}^2 \beta_i-mV_{e \phi}^2 \beta_e)}{2 r_0^2}\right]\right\} . \nonu
\eea
 The dependence of $f_i$ on $C_i$ results in an ion fluid flow,
\beq
\label{ionf}
u_{i \phi}=\frac{\int v_\phi f_i d^3 v}{n_i}=\frac{r}{r_0}V_{i \phi}, 
\eeq
with the electron fluid flow given by
\beq
\label{elecf}
u_{e\phi}=\frac{\int v_\phi f_e d^3 v}{n_e}=\frac{r}{r_0}V_{e \phi}. 
\eeq
It may be noted that the $r$-dependence of the macroscopic velocities (\ref{ionf}) and (\ref{elecf}) relates to the toroidicity; the respective flows in translational symmetric plasmas are rigid body-like. 
For the electric current density  one calculates by (9) using (\ref{Phi})
\beq
\label{cur}
j_\phi=a r \exp\left[b \Psi + c r^2\right],
\eeq
with 
$$ a=e n_{e0}^{\frac{\beta_i}{\beta_e+\beta_i}}n_{i0}^{\frac{\beta_e}{\beta_e+\beta_i}}
\frac{V_{i \phi}-V_{e \phi}}{r_0}, \ \ b=\frac{e(M V_{i \phi}-m V_{e \phi} )\beta_i \beta_e}{ r_0(\beta_i+\beta_e)},
$$
$$
c=\frac{(M V_{i \phi}^2+m V_{e \phi}^2)\beta_i\beta_e}{2 r_0^2 
(\beta_i+\beta_e)}.
$$
 Note that as expected $j_\phi$ vanishes for $V_{i \phi}=V_{e \phi}$. 
The associated Grad-Shafranov equation is put in the form
\beq
\label{grad}
\frac{\partial^2 \Psi}{\partial r^2} - \frac{1}{r}\frac{\partial \Psi}{\partial r} + \frac{\partial^2 \Psi}{\partial z^2}=-r j_\phi,
\eeq
with $j_\phi$ given explicitly by (\ref{cur}). As in the case of the static Grad-Shafranov equation (29), the exponential dependence of the RHS of (\ref{grad}) guaranties uniqueness of the solution of  the respective boundary value problem if $\Psi$ is monotonically varying from the magnetic axis to the plasma boundary. 

The pressure tensor is  given by 
 $$P_{kl}=  M \int (v_k -u_{ik})(v_l -u_{il}) f_i d^3 v + m \int (v_k -u_{ek})(v_l -u_{el}) f_ed^3 v,
 \ \ k,l=r,\phi,z,
 $$
where $\bu_i$ and $\bu_e$ are the ion and electron fluid velocities.
For the distribution functions  (\ref{iondf}) and (\ref{elecdf}) the tensor becomes diagonal and isotropic, i.e. $P_{rr}=P_{\phi\phi}=P_{zz}\equiv P$ with
\bea 
\label{pres}
P&=&\frac{n_{i0}(\beta_i+\beta_e)}{\beta_i\beta_e}
\exp\left[\frac{M V_{i \phi}\beta_i(2 e \Psi r_0+r^2 V_{i \phi}}{2 r_0^2}\right] \nonu 
& & \left\{\frac{n_{i0}}{n_{e0}}
\exp\left[\frac{2 e \Psi r_0(M V_{i \phi}\beta_i+m V_{e \phi}\beta_e)
+r^2(M V_{i \phi}^2 \beta_i-m V_{e \phi}^2 \beta_e)}
{2 r_0^2}\right]\right\}^{-\frac{\beta_i}{\beta_i+\beta_e}}. \nonu
\eea
Finally the electric filed, $\bE=-\nabla \Phi$, in general does not vanish on axis. However, equilibria with $\bE$ vanishing on axis can be constructed by the procedure of  section 4.

\section{Conclusions}
We were able to establish Grad-Shafranov-like equations starting from Vlasov equation with quasineutrality in the following two cases: i)  zero electric field on magnetic axis under the condition that the Maxwellian of the electrons is multiplied by a Heaviside step function of the angular momentum, and  ii) arbitrary electric field on axis under the condition that both electron and ion Maxwellians are multiplied by exponential functions of the angular momentum. In the former case it holds $\nabla f_i=0$ but $\nabla f_e \neq 0 $ on axis and in the latter case $\nabla f_i\neq 0$ and  $\nabla f_e \neq 0 $ on axis. The Grad-Shafranov-like equation  (29) in the former case governs static equilibria  and (\ref{grad}) in the  latter case    governs   equilibria with toroidal plasma flow.  In both cases the toroidal current density on axis remains finite. Also both equations are transcendentally nonlinear in $\Psi$ and $r$ such that for monotonically varying $\Psi$ from the magnetic axis to the plasma boundary uniqueness of the solution of the elliptic boundary value problem is guarantied. Solutions of the Eqs.(29) and (42)  however can be found only numerically which is the objective of future work. In particular, we can then locate the position of the magnetic axis. 

It is very likely that other choices of $f_i$ and $f_e$ than those made in the present study, e.g.  such that  resulting in fluid flows with stronger shear,    should require numerical integrations over the velocity space in particular as far as the current density is concerned, so that the RHS of the Grad-Shafranov-like equation is implicitly given by an integral over this space.

Finally, an advantageous feature of equilibria with vanishing electric fields on axis is that according to the results of Ref. [1] this condition  makes sure that  ${\bf E}\times{\bf B}$ drift does not occur on  axis. On the other side equilibria with finite electric fields on axis may be regarded as closer to the "actual" inductive  operation of tokamaks. Thus, a more  precise treatment of this problem  should be done  macroscopically in the framework of resistive magnetohydrodynamics\cite{tht2} and microscopically by using the  Focker-Planck  instead of Vlasov equation.   However, if   "tokamak-like" solutions of  the "ideal" Vlasov equation, associated with "ideal" distribution functions as those considered here,   can be maintained by adequate sources against collisions and turbulence, they would open up the possibility to operate the experiment with weak or even without externally induced electric fields.

\vspace{10mm}

\begin{center}
{\Large \bf Acknowledgements}
\end{center}

\vspace{10mm}

Part of this work was conducted during a visit of the author G.N.T. to the 
Max-Planck-Institut f\"{u}r Plasmaphysik, Garching. The hospitality of that 
Institute is greatly appreciated.

This work was performed within the participation of the University of Ioannina 
in the Association Euratom-Hellenic Republic, which is supported in part by the 
European Union and by the general Secretariat of Research and Technology of 
Greece. The views and opinions expressed herein do not necessarily reflect those 
of the European Commission.

\end{document}